\pdfoutput=1
\PassOptionsToPackage{table}{xcolor}
\documentclass[sigconf]{acmart}
\acmConference[MSR 2022]{MSR '22: Proceedings of the 19th International Conference on Mining Software Repositories}{May 23–24, 2022}{Pittsburgh, PA, USA}
\renewcommand\footnotetextcopyrightpermission[1]{} % removes conference venue

\usepackage{multirow}
\usepackage{dblfloatfix}

\usepackage{enumitem}
\usepackage{bigstrut}
\usepackage[tikz]{bclogo}
\usepackage[skins]{tcolorbox}
\usepackage{wrapfig}
\usepackage[flushleft]{threeparttable}
\usepackage{float}
 
\definecolor{deeppink}{HTML}{D28986} 
\definecolor{pink}{HTML}{FFABA7} 
\definecolor{lightpink}{HTML}{FFCCC9} 
\usepackage[usestackEOL]{stackengine}
\usepackage{xcolor}

\usepackage{pifont}% http://ctan.org/pkg/pifont
\newcommand{\cmark}{\ding{51}}%

\usepackage[final]{listings}
\usepackage[linesnumbered,ruled,vlined]{algorithm2e}

\def\BibTeX{{\rm B\kern-.05em{\sc i\kern-.025em b}\kern-.08em
    T\kern-.1667em\lower.7ex\hbox{E}\kern-.125emX}}

\SetCommentSty{mycommfont}

\SetKwInput{KwInput}{Input}                % Set the Input
\SetKwInput{KwOutput}{Output}              % set the Output

\SetKwFunction{FMain}{Main}
\SetKwFunction{FSum}{Sum}
\SetKwFunction{FSuba}{feature\_summarizer}
\SetKwFunction{FSubb}{cluster\_creator}
\SetKwFunction{FSubc}{bellwether\_finder}

\AtBeginDocument{%
  \providecommand\BibTeX{{%
    \normalfont B\kern-0.5em{\scshape i\kern-0.25em b}\kern-0.8em\TeX}}}

\newcommand{\IT}[1]{{\bf%
DODGE(\ifx*#1$\mathcal{E}$\else#1\fi)}}

\newcommand{\fig}[1]{Figure~\ref{fig:#1}}

\newcommand{\tbl}[1]{Table~\ref{tbl:#1}}
\newcommand{\bi}{\begin{itemize}[leftmargin=0.4cm]}
	\newcommand{\ei}{\end{itemize}}
\newcommand{\be}{\begin{enumerate}[leftmargin=0.4cm]}
	\newcommand{\ee}{\end{enumerate}}

\begin{document}
\title{Methods for Stabilizing  Models across
Large Samples of Projects  \\\ (with case studies on Predicting  Defect and Project Health)}

\author{Suvodeep Majumder}
\email{smajumd3@ncsu.edu}
\affiliation{%
\institution{North Carolina State University}
\city{Raleigh}
\country{USA}}

\author{Tianpei Xia}
\email{txia4@ncsu.edu}
\affiliation{%
\institution{North Carolina State University}
\city{Raleigh}
\country{USA}}

\author{Rahul Krishna}
\email{rkrish11@ncsu.edu}
\affiliation{%
\institution{North Carolina State University}
\city{Raleigh}
\country{USA}}

\author{Tim Menzies}
\email{timm@ieee.org}
\affiliation{%
\institution{North Carolina State University}
\city{Raleigh}
\country{USA}}

\begin{abstract}
Despite decades of research,   SE   lacks widely accepted models (that offer precise quantitative stable predictions) about  what factors most influence software quality. This paper provides a promising result showing such stable models can be generated using a new transfer learning framework called ``STABILIZER''. Given a tree of recursively clustered projects (using project meta-data), STABILIZER promotes a model upwards if it performs best in the lower clusters (stopping when the promoted model performs worse than the models seen at a lower level). 

The number of models found by STABILIZER  is minimal: one for  defect prediction (756 projects) and less than a dozen for project health (1628 projects). Hence, via STABILIZER, it is possible to find a few projects which can be used for transfer learning and make conclusions that  hold across hundreds of projects at a time.  Further, the models produced in this manner offer predictions that perform as well or better than the prior state-of-the-art. 

To the best of our knowledge, STABILIZER is order of magnitude faster than the prior state-of-the-art transfer learners which seek to find conclusion stability, and these case studies are the largest demonstration of the generalizability of quantitative predictions of project quality yet reported in the SE literature.

In order to support open science,
all our scripts and data are online at https://github.com/Anonymous633671/STABILIZER.
\end{abstract}

% \keywords{Defect Prediction, Project Health, Bellwether, Clustering}

% \keywords{Machine Learning with and for SE, Mining Software Repositories}

\maketitle

\section{Introduction}
\label{sec:intro}
The mining methods presented at the MSR conference are  excellent for building models specific to individual projects. These  models cover many diverse tasks such as the detection of code smells, code refactoring~\cite{tufano2015and, mantyla2004bad, sjoberg2012quantifying} or the two case studies explored in this paper: defect prediction~\cite{ostrand2004bugs, ghotra2015revisiting, zimm09, yang2017tlel} and software project health estimation~\cite{link2018assessing, wynn2007assessing, crowston2006assessing}. Better yet, when some specialized effect holds for one project, then our data mining methods can reveal that effect by creating a model that is exactly suited to that project. Further, when the effect changes (say, in next release), our mining methods can be run again to learn yet another model that is specialized for the new current state of the project.

But this ``specialization effect'' raises its own challenges. If every model we build is specialized to just the latest data from some current project, then it becomes hard to claim general results that hold across multiple projects. Hassan~\cite{Hassan17} warns that  managers lose trust in software analytics if the results keep changing.

% Researchers and industry practitioners use data mining algorithms to build software quality models from project data. These quality models include many diverse tasks such as the detection of code smells, code refactoring~\cite{tufano2015and, mantyla2004bad, sjoberg2012quantifying} or the two case studies explored in this paper (for more on these case studies, see \S\ref{sec:case_study}):
% \bi
%     \item Defect prediction~\cite{ostrand2004bugs, ghotra2015revisiting, zimm09, yang2017tlel} prioritize where to look in a codebase such that more bugs are found sooner.
%     \item Software project health estimation~\cite{link2018assessing, wynn2007assessing, crowston2006assessing}  incrementally monitors open-source projects (which is a very different process to classic waterfall up-front effort estimation~\cite{Kocaguneli15,boehm2000cost} that is conducted before developers start coding).  
% \ei

We raise this concern since, the usual result is that the models build via data mining for SE data are not stable across multiple projects (see~\S\ref{sec:literature} and~\cite{zimm09, Me13, menzies2011local}). When different projects result in different models, it is hard to gain insights about what affects software quality. This is troubling, as Sawyer et al.~\cite{sawyer2013bi} argue, such insights are the key driver for businesses. Generating new models each time we look at new data  exhausts users' ability to draw insight.  Table~\ref{tbl:why} lists   other problems caused by conclusion instability.

One way to  find conclusions that hold across multiple projects is the ``bellwether'' method~\cite{krishna2018bellwethers,  krishna16a, mensah18z, mensah2017stratification, mensah2017investigating}.  This method assumes  that, given a community of projects, the ``bellwether'' is a project  whose data yields the best analytic (i.e., defect prediction, effort estimation) on all others\footnote{The ``bellwether''   is a sheep with a bell on its neck that that leads rest of the flock.}. This bellwether model can be used to generate stable conclusions across that community. However, current bellwether methods require an $O(N^2)$ comparison 
 of all models build from $N$ projects. Hence, that approach is hard to scale so prior bellwether papers on every reported generality across 20 projects (or less).

\begin{table*}[!t]
  \caption{There are many reasons to seek stable general conclusions in software
          engineering. If our conclusions about best practices for SE projects keep
          changing, that will be detrimental to generality trust, insight, training, and tool
          development.}\label{tbl:why}
 { \footnotesize \begin{tabular}{|rp{.8\linewidth}|}\hline

  {\em Generality:}& Data science for SE cannot be called a ``science'' until it makes general conclusions that hold across  multiple  projects. Without  general rules for a large number of software projects, then it is   difficult to demonstrate such generality.
\\
 \rowcolor{lightpink}
{\em Trust:}&
 Hassan~\cite{Hassan17} cautions that 
managers lose faith
 in software analytics if its models keep changing
 since  the assumptions used to 
make prior policy decisions may no longer hold. \\
{\em Insight:}& Kim et al.~\cite{Kim2016}, say  that the aim of software analytics is to obtain actionable insights that help practitioners accomplish software development goals. These insights can be core deliverable~\cite{tan2016defining}, key driver for businesses ~\cite{sawyer2013bi} or user reflections~\cite{Bird:2015}. But if  new models keep being generated in new projects,  then that exhausts the ability of  users to draw insight from  new data.\\
 
 \rowcolor{lightpink}
       {\em Training:} &Another concern is what do we train novice software engineers
          or newcomers to a project?  
          If our models are not stable, then it hard to teach
          what factors  most influence software quality.\\
        {\em Tool development:}&Further to the last point---
   if we are unsure what 
          factors most influence quality, it is difficult to design and implement and deploy tools
          that can successfully improve that quality.\\\hline
          \end{tabular}}
          \end{table*}   
  
This paper proposed a novel framework called STABILIZER that works for 100s to 1000s of projects.  STABILIZER uses hierarchical clustering. At each level of the cluster hierarchy, models ``fight it out'' to select a bellwether. The winner is promoted up the hierarchy and the process repeats. Sometimes this will return one model  from the root of the hierarchy (in the case of defect prediction). And sometimes that promotion stops at some internal level  (e.g., for project health estimation, we return around a dozen models). More importantly, STABILIZER   indexes those models in its cluster tree
so, when new projects arrive, they can explore down the cluster tree in log-time to find their relevant model.
 
To assess  STABILIZER, we ask five questions:

{\bf RQ1:} {\em is STABILIZER 
   tractable?} Since one of  our primary concern with prior work was its   scalability, we are motivated to consider the computational cost of any new alternative. For $N$ projects recursively clustered into $m$ sub-groups,  STABILIZER runs in time $\Theta(m*(N/m)^2)$ (compared to $\Theta(N^2)$ for prior work).  
   
{\bf RQ2:} {\em is STABILIZER efficient?} In practice, STABILIZER's run-time increases  less-than-linearly, compared to the prior SOTA\footnote{SOTA = State of the art.} in this area~\cite{krishna2018bellwethers}, as the project community size increases.
Hence,STABILIZER scale much better than prior transfer learning methods.
For example,
in this paper, we can
show STABILIZER results for 756 defect prediction data sets
and 1628 effort estimation data sets. Those results were generated   29 to 104 times faster than   Krishna et al.'s BELLWETHER algorithm  (the prior SOTA in this area~\cite{krishna2018bellwethers}). 
    
{\bf RQ3:} {\em is STABILIZER
effective?} For all Github projects studied in this paper, the predictions made by the models generated for both case studies are better than the baseline and prior SOTA(SOTA) methods.
    
{\bf RQ4:} {\em is STABILIZER parsimonious?} We show that the number of models generated in this manner is very small (one for defect prediction and less than a dozen for project health).

{\bf RQ5:} {\em What are the generalized feature we learned?} We show that there are features which are important across a larger number of project and can be generalized across those projects.

Seeing these results, we can summarize our contributions as - 

\bi
\item A new, scalable, algorithm for transfer learning.
\item To test that new algorithm, we  offer two large case studies with 756 Github projects (for defect prediction) and 1628 Github projects (for project health estimation). To the best of our knowledge, these are the largest transfer learning experiments yet performed in the SE literature.
\item In those case studies,  our new method can generate better performing models that cover  much more projects that   seen before. 
\item We can show that  STABILIZER's new hierarchical transfer methods actually perform better traditional transfer learning methods, showing that prior transfer learning methods were not learning from enough projects.
\item We can demonstrate the    value of our new approach   via its computational complexity {\em and} extensive experimentation.
  
\ei
The rest of this paper is structured as follows. After some background  (in \S\ref{sec:literature}) we show  two cases studies: defect prediction and effort estimation. \S\ref{STABILIZER} describes our new approach. \S\ref{sec:Experimental_methods} show our experimental rig , which is used to produce the results of \S\ref{sec:results}.  

% Before beginning, we digress to make three points.

% {\bf Firstly}, here, we have demonstrated the effectiveness of STABILIZER on two domains (defect prediction and project health estimation). In future work, it would be helpful to test these methods on other domains.

% {\bf Secondly}, while we say we build general models (that hold for many projects), it is not true   that we can always generalize all projects to a single model. Indeed, as shown by the results of this paper, some domains like project health estimation require multiple models. However, even there, it is still much easier to reason about STABILIZER's dozen health models than the 1628 individual models. 

% {\bf Thirdly},  we are often asked, ``if you have (e.g.) a dozen project health models, have you won anything? How do we know which model to apply?''. We reply that our hierarchical transfer learning method is like an index that can be used to take a new project, then run it over the cluster tree to find its relevant model.

\section{Background   }
\label{sec:literature}

\subsection{Why Study Conclusion Stability?}
In this section, we justify why it is important to seek stable SE conclusions. Our concern will be with ``conclusion instability'' where conclusions change from one project to another.  For a list of issues that are raised when conclusions are unstable, see \tbl{why}.

Conclusion instability is  well documented in the research community. Zimmermann et al.~\cite{zimm09} learned defect predictors from 622 pairs of projects (project1, project2). In only 4\% of pairs, predictors from project1 worked on project2.  Menzies et al.~\cite{Me13} reported defect prediction results from 28 studies, most of which offered widely differing conclusions about what most influences defects.  

Conclusion instability is also rampant in the developer community. Multiple studies report that human beliefs in software quality may often be inconsistent and even incorrect. Devanbu et al.~\cite{devanbu2016belief, bird2011don} have conducted a case study among 564 Microsoft software developers to show that human beliefs on software quality can be quite varied and may not necessarily correspond with actual evidence within current projects. A more recent study by Shrikanth et al.~\cite{shrikanth2019assessing} also reports such variability of human beliefs. They studied ten beliefs held by software developers about defect prediction, which were initially summarized by Wan et al. in 2018 ~\cite{wan2018perceptions}. By measuring the actual support of these beliefs within the project, Shrikanth et al. found that among over 300,000 changes seen in different open-source projects, only 24\% of the projects support all ten beliefs. Curiously, beliefs held by most developers do not necessarily have the most substantial support within projects. For example, according to Shrikanth et al., a belief acknowledged by 35\% of the developers have the most support. In contrast, a belief held by 76\% of the developers is only ranked 7th out of ten beliefs. Worse still, as a project grows to mature, the beliefs tend to be weakened rather than strengthened.

Such instability prevents project managers from offering clear guidelines on many issues, including (a)~when a specific module should be inspected, (b)~when modules should be refactored,  and (c)~deciding where to focus on expensive testing procedures. Bird et al.~\cite{Bird:2015} notes that insights occur when users respond to software analytics models. Frequent model changes could exhaust users' ability for confident conclusions from new data. Further, conclusion instability   makes it hard to on-board novice engineers. It is hard to design and build appropriate tools for quality assurance activities without knowing what factors most influence the local project.

SE is conducted on ever-changing platforms using ever-changing tools by a constantly changing and evolving developer population of varying skills.
Perhaps it is   folly to assume   {\em one} model can  cover something as diverse as software engineering. But if  we cannot {\em remove} instability, perhaps another tactic is to {\em reduce} it.

Maybe a better way to look at conclusion instability is to assume that - from 1000's of projects-  there might be small number of projects. Where each can act as stable model across a smaller community. In that view,  we need tools that can find {\em and index} that small set of models that best represent the rest of the space; i.e. we need bellwethers.

% \be
%     \item From 1000's of projects,
%     \item There might be a much smaller number of models,
%     \item So we need tools that can find {\em and index} that small set of models.
% \ee 
% Note that in this approach, the indexing is important: unless we can map new projects to relevant models, then having a zoo of models will always confuse us.
What we show here is that we can reduce 1000s of projects
to just a few dozen most representative examples.
Dozens of models are still more than one, so developers and managers still need to be aware of (and debate) the aspects of their project that make it distinctive. That said, once we can summarize many projects into just a handful of models, then those debates need not be extensive. Further, when training engineers, we could present a small range of projects to them as part of their training.

\subsection{Two Case Studies}
\label{sec:case_study}
In the following, we offer a hierarchical transfer learning framework for oyr three-step process. After hierarchically clustering the data, it replaces many models (in a sub-tree) with a single best model (selected from that sub-tree). This process then recurses up the tree of clusters. Sometimes this will return one model (in the case of defect prediction). If that is not possible, it will return multiple models (e.g., for project health estimation, we return around a dozen models). More importantly, STABILIZER {\em indexes those models} in its cluster tree. When new projects arrive, they can explore down the cluster tree in log-time to find their relevant model.

This novel approach is tested on the  two case studies discussed in this section: defect prediction and project health estimation:
\bi
    \item Defect prediction~\cite{ostrand2004bugs, ghotra2015revisiting, zimm09, yang2017tlel} prioritize where to look in a codebase such that more bugs are found sooner.
    \item Software project health estimation~\cite{link2018assessing, wynn2007assessing, crowston2006assessing}  incrementally monitors open-source projects (which is a very different process to classic waterfall up-front effort estimation~\cite{Kocaguneli15,boehm2000cost} that is conducted before developers start coding).  
\ei

\subsubsection{\textbf{Why Defect Prediction?}}
\label{sec:dp}
The perceived criticality and bugginess of the code for managing resources efficiently are often associated with the QA effort. As bugs are not evenly distributed across the project~\cite{hamill2009common, koru2009investigation,  ostrand2004bugs, misirli2011ai}, it is impractical and inefficient to distribute equal effort to every component in a software system~\cite{briand1993developing}. Algorithms that measure the criticality or bugginess of software using source code (product) or project history (process) are called defect prediction models. Although such defect predictors are never 100\% correct, they can suggest where a defect might happen.

In a recent paper, Wan et al.~\cite{wan2018perceptions}, reported much industrial interest in these predictors since the alternative is much more time-consuming and expensive.  Misirili et al.~\cite{misirli2011ai} and Kim et al.~\cite{kim2011empirical} report considerable cost savings when such predictors are used in guiding industrial quality assurance processes.

As shown in \tbl{metric} defect prediction models might use {\em product} or {\em process} metrics to make predictions. Process metrics comment on ``who'' and ``how'' the code was written, while product metrics record ``what'' was written. Researchers and industry practitioners have tried many different ways to identify which features are important and why. However, there is little agreement between them. Zimmermann et al.~\cite{zimmermann2007predicting} recommended complexity-based product metrics, Zhou et al.~\cite{zhou2010ability} suggest size-based metrics. While Matsumoto et al.~\cite{matsumoto2010analysis}, and Nagappan et al.~\cite{nagappan2010change} recommend developer-related metric and change bursts metrics, respectively. We use \tbl{metric} metrics for both theoretical and pragmatic reasons. Our metrics include the process metrics endorsed by Devanbu et al. at ICSE'13~\cite{rahman13}. 

% Also, tools exist for collection the \tbl{metric} metrics across hundreds to thousands of Github projects such as the Commit\_Guru tool~\cite{rosen2015commit} used in this study, discussed below.

\subsubsection{\textbf{Why Project Health?}}
\label{sec:ph}

Open-source software development is becoming prominent in the overall software engineering landscape. As the community matures, they become more structured with organizing foundations. Some prominent software organizations like Apache Foundation and Linux Foundation host hundreds of popular projects~\cite{apacheprojects, linuxprojects}.  Stakeholders of these projects make critical decisions about the future of these projects based on project status. The metrics of project health conditions are needed when they estimate the projects.  On the other hand, future customers thinking about using the open-source project in their product are more likely to invest and participate in ``healthy'' projects. 

In the past decade, many researchers~\cite{jarczyk2018surgical, kikas2016using, chen2014predicting, bidoki2018cross, xia2020predicting} have tried to look at project health from a different perspective. For example, Jansen et al.~\cite{jansen2014measuring} mentioned project could be estimated by the level of  productivity, robustness, and niche of creation in the project history. Kikas et al.~\cite{kikas2016using} reported a relationship between the dynamic and contextual features of a project and issue close time.  Wang et al.~\cite{wang2018will} and Bao et al.~\cite{bao2019large} proposed different predicting models to find potential long-term contributors. Jarczyk et al.~\cite{jarczyk2018surgical} use generalized linear models for the prediction of issue closure rate. Based on multiple features (i.e., stars, commits, issues closed), they find that larger teams with more project members have lower issue closure rates than smaller teams. At the same time, increased work centralization improves issue closure rates.

From January to March 2021, Xia~\cite{xia21_a} conducted email interviews with 116 subject matter experts from 68 open-source projects. One of the questions asked was ``what kind of features are the most useful to predict for?''. In those answers, certain  features were marked as the most  important to that community: see \tbl{project_health_data}. Note the ``Predict?'' column of that table: in our experiments, we take each prediction goal, one at a time, then try to predict for it using all the other metrics for six months into the future.

\begin{table}[]
\caption{List of metrics used in this study for defect prediction case study.}
\scriptsize
\centering
\begin{tabular}{l|l|l|l}
\rowcolor[HTML]{C0C0C0} 
Metric & Description                                              & Metric & Description                                    \\ \hline
adev   & Active Dev Count                                         & nadev  & Neighbor’s Active Dev Count                    \\
age    & Interval between last, current change                 & ncomm  & Neighbor’s Commit Count                        \\
ddev   & Distinct Dev Count                                       & nd     & Number of Modified Directories                 \\
sctr   & Distribution of modified code                            & nddev  & Neighbor’s Distinct Dev Count                  \\
exp    & Experience of the committer                              & ns     & Number of Modified Subsystems                  \\
la     & Lines of code added                                      & nuc    & Number of unique changes to files \\
ld     & Lines of code deleted                                    & own    & Owner’s Contributed Lines                      \\
lt     & Lines of code before the change                          & sexp   & Developer experience on a subsystem            \\
minor  & Minor Contributor Count                                  & rexp   & Recent developer experience                   
\end{tabular}
% \begin{tabular}{r@{~:~}l}
% adev     &  Active Dev Count                                 \\  
% age      &  Average interval between the last and the current change \\  
% ddev     &  Distinct Dev Count                               \\  
% sctr  &  Distribution of modified code across each file      \\
% exp      &  Experience of the committer                      \\  
% la       &  Lines of code added                              \\  
% ld       &  Lines of code deleted                            \\  
% lt       &  Lines of code in a file before the change        \\  
% minor    &  Minor Contributor Count                          \\  
% nadev    &  Neighbor’s Active Dev Count                      \\  
% ncomm    &  Neighbor’s Commit Count                          \\  
% nd       &  Number of Modified Directories                   \\  
% nddev    &  Neighbor’s Distinct Dev Count                    \\  
% ns       &  Number of Modified Subsystems                    \\ 
% nuc      &  Number of unique changes to the modified files   \\  
% own      &  Owner’s Contributed Lines                        \\  
% sexp     &  Developer experience on a subsystem              \\
% rexp     & Recent developer experience                       \\
% \end{tabular}
\label{tbl:metric}
% \vspace{-0.6cm}
% \end{table}
\bigskip
% \begin{table}[]
\caption{List of metrics used in this study for project health estimation case study.}
\scriptsize
\centering
\begin{tabular}{l|lc}
\rowcolor[HTML]{C0C0C0} 
\begin{tabular}[c]{@{}l@{}}Metric \\ Category\end{tabular}                                                   & Metric Name                                                    & Predict?             \\ \hline
                                                                                                             & \multicolumn{1}{l|}{MC: monthly number of commits}              & \cmark                     \\
\multirow{-2}{*}{Activeness}                                                                                   & \multicolumn{1}{l|}{MAC: monthly number of active contributors} & \cmark                     \\ \hline
                                                                                                             & \multicolumn{1}{l|}{MOP: monthly number of open PRs}            & \cmark                     \\
                                                                                                             & \multicolumn{1}{l|}{MCP :monthly number of closed PRs}          & \cmark                     \\
                                                                                                             & \multicolumn{1}{l|}{MMP: monthly number of merged PRs}          &                      \\
                                                                                                             & \multicolumn{1}{l|}{MPM: monthly number of PR mergers}          &                      \\
\multirow{-5}{*}{\begin{tabular}[c]{@{}l@{}}Collaboration\end{tabular}} & \multicolumn{1}{l|}{MPC: monthly number of PR comments}         &                      \\ \hline
                                                                                                             & \multicolumn{1}{l|}{MOI: monthly number of open issues}         & \cmark                     \\
                                                                                                             & \multicolumn{1}{l|}{MCI: monthly number of closed issues}       & \cmark                     \\
                                                                                                             & \multicolumn{1}{l|}{MIC: monthly number of issue comments}      &                      \\
\multirow{-3}{*}{\begin{tabular}[c]{@{}l@{}}Enhancement\end{tabular}} & \multicolumn{1}{l|}{}                                          & \multicolumn{1}{l}{} 
\\ \hline
                                                                                                             & \multicolumn{1}{l|}{MS: monthly increased number of stars}      & \cmark                     \\
                                                                                                             & \multicolumn{1}{l|}{MF: monthly increased number of forks}      &                      \\
\multirow{-3}{*}{\begin{tabular}[c]{@{}l@{}}Popularity\end{tabular}}                              & \multicolumn{1}{l|}{MW:monthly increased number of watchers}   &                     
\end{tabular}
\label{tbl:project_health_data}
\vspace{-0.6cm}
\end{table}

\subsection{\textbf{Transfer Learning}}
\label{tion:how1} 
 
From a formal perspective, STABILIZER is a {\em   homogeneous, similarity-based, bellwether, transfer learning} algorithm (and this section explains all the terms in {\em italics}. 

The art of moving data and knowledge  from one project or another is  called {\em transfer learning}~\cite{pan2009survey}. When there is insufficient data to apply data miners, transfer learning can be used to transfer knowledge from other source projects to the target project. Clark and Madachy~\cite{clark15} in their study showed developers working in an uncommon area often benefit from transferring knowledge from more common areas.

% of 65 software under-development by the US Defense Department in 2015 showed developers working in an uncommon area often benefit from transferring knowledge from more common areas.

While transfer learning is widely studied and used in the software engineering domain, we warn that work mostly tries to move lessons learned from a single source project to a single target project, and these source projects changes for new target project. Hence, that research does not meet the goals of this paper (conclusion stability across 100s of projects). As discussed in the next section, additional engineering is required to convert current transfer learning tools for conclusion stability.

Transfer learning can be broadly categorized into two variants based on the similarity of features between source and target projects. Heterogeneous transfer learning is where the source and target data contains different attributes~\cite{jing2015heterogeneous, he2014towards, nam2017heterogeneous, cheng2016heterogeneous, yu2017feature}. {\em Homogeneous}  transfer learning is where the source and target data contains the same attributes~\cite{ma2012transfer, zimm09, turhan2009relative, krishna2018bellwethers}. Our reproduction package for STABILIZER (listed on page1) collects the same set of attributes for all projects in both (a)~defect prediction or (b)~project health estimation case study. Hence, STABILIZER is a {\em homogeneous} transfer learner.

Another way to divide transfer learning is the approach that it follows. One is dimensional transformation methods, which manipulate the raw source data until it matches the target. Ma et al.~\cite{Ma2012} introduced dimensional transformation using transfer naive Bayes (TNB). Since then there are many such algorithms have been proposed such as TCA~\cite{Nam13}, TCA+~\cite{Nam2015}, TPTL~\cite{liu2019two}, balanced distribution~\cite{xu2019cross}. The second is {\em Similarity-Based} approach. Like Burak filter~\cite{turhan09} or combining domain knowledge with automatic processing of features~\cite{zhang2016towards}. Our current implementation for STABILIZER is based around the BIRCH clustering algorithm~\cite{zhang1996birch} that groups together projects with similar  median values for their features. Hence, STABILIZER is a similarity-based algorithm.

% Here in this study, we have used the TPTL method proposed by Liu et al.~\cite{liu2019two} as the SOTA (SOTA) method\footnote{This paper out-performs many of the previous methods (i.e., TCA+~\cite{Nam2015}, TDS~\cite{herbold2013training}, LT~\cite{cruz2009towards}, Dycom~\cite{minku2015make}) for homogeneous transfer learning.}. Note that we could only apply TPTL to defect prediction since there is no prior work on transfer learning for project health\footnote{ The nearest work we found was for some classical waterfall-based effort estimation from Kocaguneli1 et al.~\cite{Kocaguneli15}. Waterfall estimation is done prior to developers starting to code. Hence it is very different from the incremental monitoring process used in project health. Also, waterfall estimation predicts for ``total developer effort'' which, as seen in \tbl{project_health_data}, is {\em not} a metric that was not found relevant to the open-source software practitioners surveyed by Xia~\cite{xia21_a}.}. 

Another useful transfer learning approach is Krishna et al.'s {\em bellwether} method~\cite{krishna2017simpler,krishna16}. According to Krishna et al., within a community of software projects, there is one exemplary project called the bellwether, which can define predictors for the others. This effect is called {\em the bellwether effect}. They exploit this Bellwether effect in their {\em  bellwether method}  that searches for such an exemplar bellwether project to construct a transfer learner with it\footnote{According to the Oxford English Dictionary, the bellwether is the leading sheep of a flock, with a bell on its neck, that all other sheep follow.}.

The genesis of this paper was the realization that Krishna et al.'s methods had trouble scaling to the 756 and 1628 projects used in this study. Krishna et al. found their bellwether via an $\Theta(N^2)$ search through all pairs of projects. We found we could do much better than that by our STABILIZER framework. That said, the origins of this paper are clearly the methods of Krishna et al. Hence it is appropriate to say that STABILIZER is a bellwether method.

\section{  STABILIZER: The Details}
\label{STABILIZER}
% Initially, we thought that it would be enough to just run the BIRCH clustering algorithm~\cite{zhang1996birch} prior to running bellwether~\cite{krishna2017simpler}. However, as shown by Algorithm~1, we had to implement a large number of additional engineering details before anything worked, since STABILIZER consists of many sub-routines:

The STABILIZER framework consists of many sub-routines:
\be
    \item Summarize the projects via {\bf feature summarization} (see \S\ref{sec:fx}) with an average-case complexity of $\Theta(N)$.
    \item Group all our data into sets of similar projects in a hierarchical structure.  This step requires a {\bf hierarchical clustering} algorithm called BIRCH~\cite{zhang1996birch} (see \S\ref{sec:hc}) with  a complexity of $\Theta(N)$ where a feature vector from step one represents each project.
    \item Select the best source project (bellwether) at each cluster at the leaves of that tree. This step needs a {\bf data mining} algorithm to generate models (see \S\ref{sec:dm}) and  a comparison method to {\bf select the best model} (see \S\ref{sec:best}). Here assuming an average-case scenario where $N$ projects are divided into $m$ clusters, the average-case complexity is $\Theta(m*(N/m)^2)$.
    \item  Select the best source project at non-leaf levels by pushing bellwether from child nodes. Each super-group only has projects selected as bellwether from step three then steps three, four are repeated, recursively until root node. This requires a  complexity of $\Theta(m^2)$.
    \item Use the bellwethers to predict or estimate for new projects.
\ee
\begin{algorithm}[!t]
\scriptsize
\DontPrintSemicolon
    \KwInput{Datasets, Training\_Projects, Test\_Projects}
    \KwOutput{Test\_Results, Feature\_Importance}
    \SetKwProg{Fn}{Def}{:}{}
        \Fn{\FSuba{$Datasets$}}
        {
            feature\_vecs = []\;
            \For{dataset in Datasets}
            {
                feature\_vecs.append(dataset.summarize())\;
            }
            \KwRet feature\_vecs\;
        }
        \SetKwProg{Fn}{Def}{:}{}
        \Fn{\FSubb{$feature\_vecs$}}
        {
            CF\_Tree = BIRCH(feature\_vecs)\;
            \KwRet CF\_Tree\;
        }
        \SetKwProg{Fn}{Def}{:}{}
        \Fn{\FSubc{$projects$}}
        {
            performance = []\;
            \For{s\_project in projects}
            {
                clf = model(s\_project)\;
                 \For{t\_project in projects}
                {
                    result = clf.score(t\_project)\;
                    performance[s\_project] = result\;
                }
            }
            bellwether = best(performance)\;
            \KwRet bellwether\;
        }
        
    feature\_vecs = feature\_summarizer(Datasets)\;
    CF\_Tree = cluster\_creator(feature\_vecs)\;
    bellwethers = []\;
    \For{level in CF\_Tree.levels}
    {
        \If{level == CF\_Tree.max\_levels}
        {
            \For{cluster in CF\_Tree[level].clusters}
            {
                bellwether = bellwether\_finder(cluster.projects)\;
                bellwethers[level][cluster] = bellwether\;
            }
        }
        \Else
        {
            \For{cluster in CF\_Tree[level].clusters}
            {
                child\_clusters = CF\_Tree[level][cluster].child\;
                bellwether\_projects = bellwethers[level][child\_clusters]\;
                bellwether = bellwether\_finder(bellwether\_projects)\;
                bellwethers[level][cluster] = bellwether\;
            }
        }
    }

\caption{Pseudocode of STABILIZER}
\vspace{-0.2cm}
\label{Algo}
\end{algorithm}
% Note assuming in an average-case clustering algorithm divides the $N$ projects into $m$ clusters at the leaf level, then the  average-case complexity of the  algorithm~\ref{Algo} (which we call STABILIZER) can be represented as:
% \begin{equation}
% \label{eq:STABILIZER}
%     {STABILIZER\; complexity (Average\; case)} =  \Theta(m*(N/m)^2)
% \end{equation}
% Comparing the computational cost of Equation~\ref{eq:STABILIZER} with Equation~\ref{eq:bellwether}, we can say the $\Theta(m*(N/m)^2)$ analysis is inherently more scalable than the $\Theta(N^2)$ analysis required by standard bellwether as m increases.
The rest of this section documents our design choices made
for those sub-routines:

\subsection{Feature Summarization}
\label{sec:fx}
Prior to anything else, we must summarize our  project into a single vector to be used in the clustering algorithm.  We follow the direction of Herbold et al.~\cite{herbold2013training} and Liu et al.~\cite{liu2019two} as they have shown if data distributions between the source and target projects are close, a cross-project defect prediction (CPDP) model can achieve better prediction performance. Hence, for both case studies, we summarize  using the median values of the dataset.

\subsection{Hierarchical Clustering}
\label{sec:hc}
Using the feature vectors for each project, we apply a hierarchical clustering algorithm to communities of similar projects. We use the Balanced Iterative Reducing and Clustering using Hierarchies  (BIRCH) algorithm  from the scikit.learn~\cite{scikit-learn} package. BIRCH is suitable for large datasets that might contain spurious outliers~\cite{zhang1996birch} and can incrementally and dynamically cluster incoming, multi-dimensional data in an attempt to maintain the best quality clustering. BIRCH also can identify data points that are not part of the underlying pattern (so it can effectively identifying and avoid outliers). For this, we used defaults proposed by~\cite{zhang1996birch}, a branching factor of 20, and  the ``new cluster creation'' threshold  of  0.5.

\subsection{Data Mining}
\label{sec:dm}
The new bellwether analysis described above requires a working data miner framework. The two different case studies in this paper need a slightly different framework as defect prediction is a classification task, while project health estimation is a regression task. The main requirements for that data miner framework are:

(a)~\textbf{feature selector} to prune any unnecessary features (both classification and regression tasks).

(b)~\textbf{class balancing technique} to balance the dataset (for classification tasks).

(c)~\textbf{hyper-parameter optimizer} to tune the performance of the models (both classification and regression tasks).

(d)~\textbf{ML learner} to generate models (both classification and regression tasks) and find answers to ``what did we learn?''

To fulfill these requirements, (a)~for feature selector, we have used Hall's \textit{CFS feature selector}~\cite{hall1999correlation}.  We found that without CFS, our recalls were low. Also, extensive studies have found that CFS is more useful than many other feature subset selection methods such as  PCA, InfoGain, or RELIEF~\cite{hall1999correlation, challagulla2008empirical, arar2015software, lee2016developer, rodriguez2007attribute, gao2015combining, hosni2017investigating}. (b)~for class balancing technique, \textit{SMOTE}~\cite{Chawla2002}. We used SMOTE since it is a widely used~\cite{tan2015online, wang2013using, sun2012using, khoshgoftaar2010attribute} class imbalance correction algorithm. We use SMOTE  on the training data\footnote{While it is useful to artificially boost the number of target examples in the training data~\cite{Chawla2002, Pelayo2007, mensah2017investigating}, it is a methodological error to also change the distributions in   the test data~\cite{agrawal17}. Hence, for our work, we take care to {\em only} re-sample the training data.} (c)~for hyper-parameter optimization; we used \textit{Differential Evolution (DE)}~\cite{storn1997differential}.  We use DE since prior work found it fast and comparatively more effective than grid search for other kinds of software analytic problems ~\cite{onan2016multiobjective, tantithamthavorn2018impact, fu2016differential, Fu2016TuningFS, zhang2018cost}. (d)~for the learner, we have used \textit{Random Forest}~\cite{breiman2001random}, \textit{Decision Tree}~\cite{breiman1984classification} and  we used these as it is widely used~\cite{ghotra2015revisiting, tantithamthavorn2018impact, yang2017tlel} in the software engineering domain as the learner and shown to be effective in selecting important features~\cite{saeys2008robust, uddin2015guided}. 

For both \textbf{defect prediction} and \textbf{project health estimation} case study, on the original data, we apply \textit{CFS}~\cite{hall1999correlation}. Next, for defect prediction, we use the  \textit{SMOTE}~\cite{Chawla2002} to balance the classes. And finally, we used \textit{Random Forest} as the learner for defect prediction and \textit{Decision Tree Regressor} with DE for project health estimation.

% \begin{figure*}[]
%     \centering
%     \includegraphics[width=\linewidth]{STABILIZER.pdf}
%     \caption{Experimental rig for this paper.  In this rig, bellwethers are learned and tested on {\em separate } projects. Within the test set (denoted ``test1'', above), the data is further divided into ``train2, test2''. To assess the bellwether found from ``train1'' against local learning, data from each project in test1 is divided into ``train2,test2'' then (a)~local models are learned from   ``train2''; after which time, (b)~the local models from ``train2'' and the bellwether model from ``train1'' are both applied to the same data from ``test2''. Note that this process is repeated ten times, with different random number seeds, to generate ten different sets of ``train1, train2, test2''.}
%     \label{fig:STABILIZER}
% \end{figure*}

\subsection{Evaluation Criteria}
\label{sec:Measures}
We used five evaluation criteria to evaluate our defect prediction models (two effort aware and three traditional). Suppose we have a dataset with M changes and N defects. After inspecting 20\% LOC, we inspected $m\%20$ changes and found $n\%20$ defects ($m$ changes and $n$ defects when inspected 100\% LOC). Also, when we find the first defective change, we have inspected k changes. Using this data, we can define two effort aware evaluation criteria as (a)~\textbf{Popt20:}  is the proportion of changes inspected by reading 20\% of the code, it is computed as $m\%20/M$, and (b)~\textbf{IFA:}  is the number of initial false alarms encountered before we identify the first defective change. As for the three traditional measures, they can be explained as (a)~\textbf{Recall:} is the proportion of inspected defective changes among all the actual defective changes, it is computed as $n/N$, (b)~\textbf{Precision:} is the proportion of inspected defective changes among all the inspected changes, it is computed as $n/m$, and (c)~\textbf{False alarm:} is the proportion between the non-defective changes among all the predicted defective changes.

Now to evaluate the models build to estimate project health (regression task), we use Magnitude of the Relative Error (MRE)~\cite{shepperd2012evaluating, sarro2016multi, xia2020predicting}. \textbf{MRE} is the absolute residual (AR) as a ratio of actual value. It is calculated as $\mathit{MRE} = |\mathit{predicted} - \mathit{actual}|/\mathit{actual}$.

\subsection{Select the Best Model}
\label{sec:best}
The previous section listed numerous evaluation criteria (goal) that could be used to guide the learning. Combining numerous evaluation criteria is often implemented as a multi-objective optimization predicate, where one model is better than another if it satisfies a ``domination predicate''. We use the Zitler indicator dominance predictor~\cite{zit02} to select our bellwether (since this is known to select better models for multi-goal optimization in SE~\cite{Sayyad:2013, Sayyad:2013:SPL})

% \footnote{An alternative to the Zitler indicator is   ``boolean domination'', which says one thing is better than another if it is no worse on any criteria and better on at least one criterion. We prefer Equation~\ref{eq:cdom} to boolean domination since we have a  three-goal optimization problem, and it is known that boolean domination often  fails for two or more goals~\cite{Wagner:2007, Sayyad:2013}.}.

% This predicate favors model $y$ over $x$  model if $x$ ``losses'' most:
% {\small
% \begin{equation}
% \label{eq:cdom}
%     \begin{array}{rcl}
%         \textit{worse}(x,y)& =& \textit{loss}(x,y) > \textit{loss}(y,x)\\
%         \textit{loss}(x,y)& = &\sum_j^n -e^{\Delta(j,x,y,n)}/n\\
%         \Delta(j,x,y,n) & = & w_j(o_{j,x}  - o_{j,y})/n
%     \end{array}
% \end{equation}}
% where  ``$n$'' is the number of objectives (for us, $n=5$) and $w_j\in \{-1,1\}$ depending on whether we seek to maximize goal $x_j$.  

When choosing a model based on multiple goals (defect prediction case study) using Zitler indicator dominance predictor. Here model performance can be scored, and the best model can be chosen by researchers based on their choice of metrics and their importance to them. In this experiment, we score model performance according to the goals mentioned in \S\ref{sec:Measures}. When assessing models, we  chose the best model that: (a)~Maximize Recall, Precision, and Popt20, and (b)~Minimizes false alarm, IFA, and MRE.

\section{Experimental Methods}
\label{sec:Experimental_methods}

This section describes the rig used to evaluate the design choices made in the previous section.
\subsection{Experimental Setup}
\label{sec:Experimental}

% \fig{STABILIZER} illustrates our experimental rig. The following process was repeated ten times, with different random seeds used each time.

All projects are divided randomly into two groups train\_1 and test\_1, with a 90:10 split. The projects in train\_1 were used to find bellwethers. Data from each project in test\_1 was then divided into train\_2 and test\_2 (using a 2:1 split). The defect prediction framework and project health framework (as described in \S\ref{sec:dm}) are used to build models for the two case studies along with chosen baselines and SOTA learners (as described in \S\ref{sec:mylearners}). Finally, both models were then applied to the test\_2 data to measure performance. The above mentioned process was repeated ten times, with different random seeds used each time.

\subsection{Data Collection}
\label{sec:data}
Github stores millions of projects. Many of these are trivially very small, not maintained, or not about software development projects. To filter projects, we used the standard Github ``sanity checks'' recommended in the literature ~\cite{perils, curating}:
 
\bi
    \item {\textit{{Collaboration}}: must have at least one pull request.}
    \item {\textit{{Commits}}: must contain more than 20 commits.}
    \item {\textit{{Duration}}: must contain  at least 50 weeks of development activity.}
    \item {\textit{{Issues}}: must contain more than ten issues.}
    \item {\textit{{Personal}}: must have at least ten contributors.}
    \item {\textit{{Software Development}}: can only be a placeholder for software development source code.}
    \item {\textit{{Defective Commits}}: must have at least ten defective commits with defects on Java files.}
    \item {\textit{{Forked Project}}:    not   a forked project from   original repository.}
\ei

Using these sanity checks from a community of 5000 Github projects,  we selected 756 Github Java projects for defect prediction case study. For each project, the defect prediction data was collected in the following steps:
\be
    \item Using a modified version\footnote{Commit\_Guru summarizes each commit by taking the median value of all files modified in the commit. We store the file-specific information before that step. We also added neighbor based metric calculation with the original Commit\_Guru metrics.} of Commit\_Guru~\cite{rosen2015commit} code, we collect 21 file-level process metrics, as shown in Table~\ref{tbl:metric}.
    \item We use Commit\_Guru~\cite{rosen2015commit} code to identify commits with bugs. This process involves identifying commits that were used to fix some bugs using a keyword-based search. Such as - 
    \begin{center}
        \textit{(bug $\mid$ fix $\mid$ error $\mid$ issue $\mid$ crash $\mid$ problem $\mid$ fail $\mid$ defect $\mid$ patch)}
    \end{center}
\ee

For the project health estimation case study, we use the dataset collected by Xia et al.~\cite{xia2020predicting}. 1628 projects collected from their study passed our sanity checks. This dataset records a set of developing activities in monthly counts. Following the direction of many researchers in project health~~\cite{wahyudin2007monitoring, jansen2014measuring, manikas2013reviewing, link2018assessing,wynn2007assessing, crowston2006assessing}, this dataset contains 13 features in four different categories are mentioned in Table~\ref{tbl:project_health_data}. Based on the best of existing industrial development practices, seven features marked with a tick(\cmark) in Table~\ref{tbl:project_health_data} are used as health indicators for a project from different perspectives~\cite{xia2020predicting}.

\subsection{Learners}
\label{sec:mylearners}
In this study, we compare our models with reference, baseline, and SOTA(SOTA) learners for file-level defect prediction:  

\textbf{(a)~Self:} individual model is built using training data (train\_2), which we use as the reference model\footnote{Note: here, a reference model refers to the best prediction performance  that can be achieved for a test dataset.} in this study. 

\textbf{(b)~Global:} a single model is built by pooling all data from all projects from train\_1.

\textbf{(c)~Bellwether0:} this model is built using the project data from train\_1 returned by the  $\Theta(N^2)$ bellwether method proposed by Krishna et al.~\cite{krishna16a}.

\textbf{(d)~STABILIZER(i):} these are the models built using the bellwether projects returned by ``STABILIZER'' at different levels. Here ``i'' denotes the level at which we are trying to find a bellwether. Here except at level(0), which is the root node, all other levels will return multiple bellwether\footnote{Each level will return bellwethers equivalent to the number of clusters at that level. I.e., if  at level(2) in a cluster tree we have 50 leaf clusters, then STABILIZER(2) will return 50 bellwethers.}.

Also, in the specific case of defect prediction, we ran 

\textbf{(e)~TPTL:} as this paper is fundamentally related to the literature on transfer learning and cross-project defect prediction, we use a prior SOTA transfer learning framework proposed by Liu et al.~\cite{liu2019two} as baseline. TPTL\footnote{This paper out-performs many of the previous methods (i.e., TCA+~\cite{Nam2015}, TDS~\cite{herbold2013training}, LT~\cite{cruz2009towards}, Dycom~\cite{minku2015make}) for homogeneous transfer learning.} automatically chooses two source projects based on estimated highest values of F1-score and Popt20 and then leverages TCA+~\cite{Nam2015} to build two prediction models based on the two selected projects and combine their prediction results.  For more notes on TPTL and TCA+, please recall \S\ref{tion:how1}.

Note that TPTL was not used for project health estimation\footnote{ The nearest work we found was for some classical waterfall-based effort estimation from Kocaguneli1 et al.~\cite{Kocaguneli15}. Waterfall estimation is done prior to developers starting to code. Hence it is very different from the incremental monitoring process used in project health. Also, waterfall estimation predicts for ``total developer effort'' which, as seen in \tbl{project_health_data}, is {\em not} a metric that was not found relevant to the open-source software practitioners surveyed by Xia~\cite{xia21_a}.}.  since: it assumes that {\em classification} is being used as the learner and here, we use a {\em regression} algorithm to obtain a numeric prediction.

\subsection{Statistical Tests}
\label{eval}
When comparing the results of different models in this study, we used a statistical significance test and an effect size test:
\bi
    \item The significance test is useful for detecting if two populations differ merely by random noise. 
    \item The effect sizes are useful for checking that two populations differ by more than just a trivial amount.
\ei
Here,  we use the Scott-Knott test~\cite{mittas2013ranking, ghotra2015revisiting} which includes both significance and effect size test. This technique recursively bi-clusters a sorted set of numbers. If any two clusters are statistically indistinguishable, Scott-Knott assigns both the same ``rank''. These ranks have a different interpretation, depending on whether we seek to minimize or maximize those numbers. For our purposes:
\bi
    \item Rank 1 is {\em worst} for recall, precision, Popt20 since we want to {\em maximize} these numbers.
    \item Rank 1 is {\em best} for false alarm, IFA and MRE since we want to {\em minimize} those.
\ei

\section{Results}
\label{sec:results}
This section explores the research questions in detail:
% \bi
% \item RQ1: Is STABILIZER tractable?
% \item RQ2: Is STABILIZER efficient?
% \item RQ3: Is STABILIZER effective?
% \item RQ4: Is STABILIZER parsimonious (only generates  a few models)?
% \ei
\subsection{RQ1: Is STABILIZER tractable?}
\label{sec:rq1}
As proposed by Krishna et al., conventional bellwether compares all projects (N projects) in a given community with all others to build defect prediction models. This operation is quite expensive with, an average computational complexity of $ \Theta(N^2) $. The new bellwether method ``STABILIZER'' proposed in this paper uses hierarchical clustering to group similar projects. STABILIZER finds bellwether at leaf level clusters (m clusters) and pushes up bellwethers to parent nodes for comparisons. Algorithm~\ref{Algo} shows the pseudocode of the ``STABILIZER'' algorithm, and using this, we can calculate the complexity of ``STABILIZER''. Assuming the BIRCH clustered N projects into m clusters at leaf level (for an average-case, we are assuming each cluster have an equal number of projects), the complexity can be calculated as:

{\small
\begin{equation}
\label{eq:algo}
    \begin{array}{lll}
        {\mathit feature\_summarizer(line 18)}& : &  \Theta(N) \\
        {\mathit cluster\_creator(line 19)}& : & \Theta(N)\\
        {\mathit bellwether\_finder(line 23-25)}& : & \Theta(m*(N/m)^2)\\
        {\mathit bellwether\_finder(line 27-31)}& : & \Theta(m^2)\\
        {\mathit STABILIZER}& : & \Theta(N) + \Theta(m^2) + \\ 
                          &   & \Theta(m*(N/m)^2) + \Theta(m^2) = \\
                          &   & \Theta(m*(N/m)^2)\\
        % \textit{$STABILIZER$}& : & 1 + 2 + 3 + 4 \\ 
        %                   &   & \Theta(m*(N/m)^2)\\
        
    \end{array}
\end{equation}}

As shown in equation~\ref{eq:algo}, this process reduces the  computational complexity to $\Theta(m*(N/m)^2)$ on an average-case. Hence, by the definitions of Garey and Johnson~\cite{garey90}, STABILIZER is tractable.

\begin{figure}[!b]
    \centering
    \includegraphics[width=\linewidth]{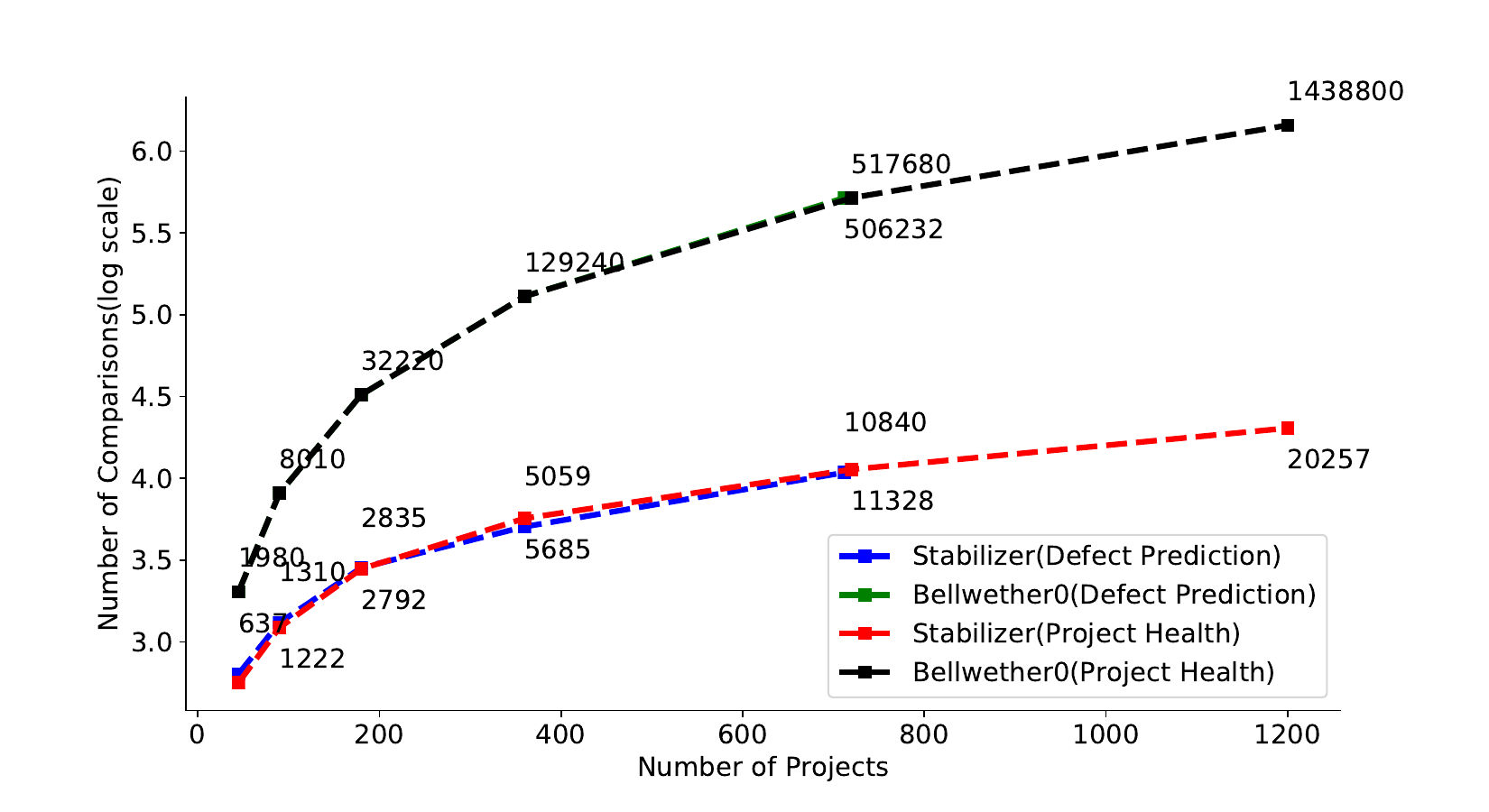}
    \caption{RQ2 results: Number of comparisons.}
    \label{fig:compare}
    \vspace{-.4cm}
\end{figure}

\begin{figure}[!b]
    \centering
    \includegraphics[width=\linewidth]{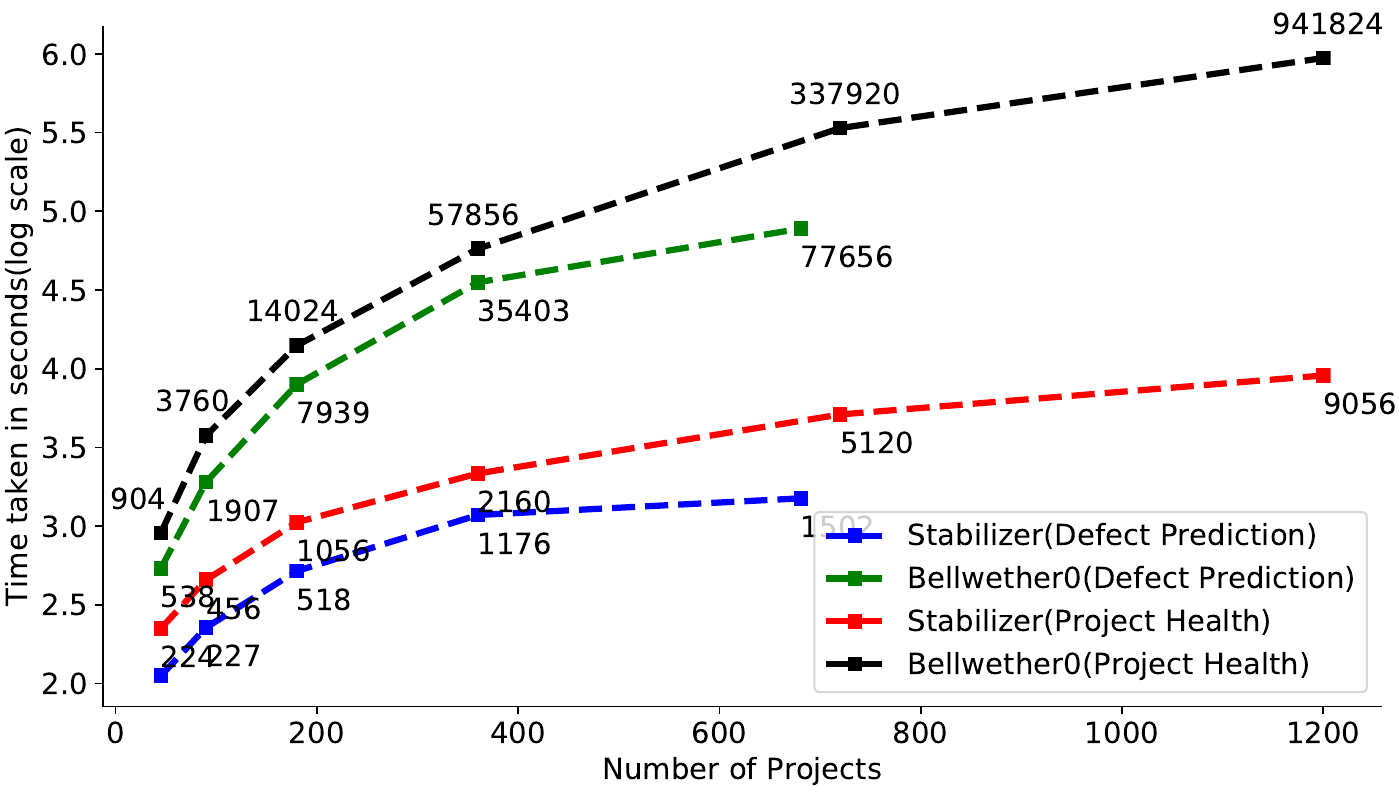}
    \caption{RQ2 results: Median run-time (secs).}
    \label{fig:compare_time}
    \vspace{-.6cm}
\end{figure}

\subsection{RQ2: Is STABILIZER efficient?}
\label{sec:rq2}
To assess the speed of STABILIZER and compare it against the conventional bellwether, we ran both defect prediction and project health estimation case study on the STABILIZER framework on a sixteen core machine running at 2.3GHz with 32GB of RAM to measure their run time. We also measure the number of comparisons performed in conventional bellwether and our proposed model on both frameworks.

\fig{compare} shows the median (over multiple runs) number of comparisons required for finding a bellwether project using conventional bellwether vs STABILIZER for different community sizes ranging from 45 to 1200 projects for both case studies (for defect prediction maximum was 711 projects). The {\color{green}\textbf{Green}} and the {\color{black}\textbf{Black}} lines show the number of comparisons required for conventional bellwether on the defect prediction and project health estimation case study, respectively. Simultaneously, the {\color{blue}\textbf{Blue}} and the {\color{red}\textbf{Red}} line show the number of comparisons for our STABILIZER algorithm. Here the y-axis is plotted in a log scale while the numbers shown are the actual number of comparisons performed. 

Similarly, \fig{compare_time} shows the median run time required for finding a bellwether.  We can see from \fig{compare} and \fig{compare_time}, with increasing community size, the number of comparisons and run-time time increases rapidly for the conventional bellwether method. In contrast, STABILIZER requires a relatively small number of comparisons and run-time (it increases less than linearly). 

% while with STABILIZER the number of comparisons required is relatively small (it increases less than linearly). 

In summary, STABILIZER
will scale to a large space of projects.
Its run-time increases  less-than-linearly as the project community size increases. STABILIZER is very efficient, requiring less than an hour to terminate for both case studies (756 and 1628 Github projects for defect prediction and project health estimate). Those run-times are 29 to 104 times faster than those seen in Krishna et al.'s BELLWETHER algorithm  (the prior SOTA in this area~\cite{krishna2018bellwethers}).

\subsection{RQ3: Is STABILIZER effective?}
\label{sec:rq3}

The speed improvements reported in {\bf RQ2} are only useful if this faster method can also deliver adequate predictions (i.e., predictions that are not worse than those generated by baselines and SOTA).

\begin{figure}
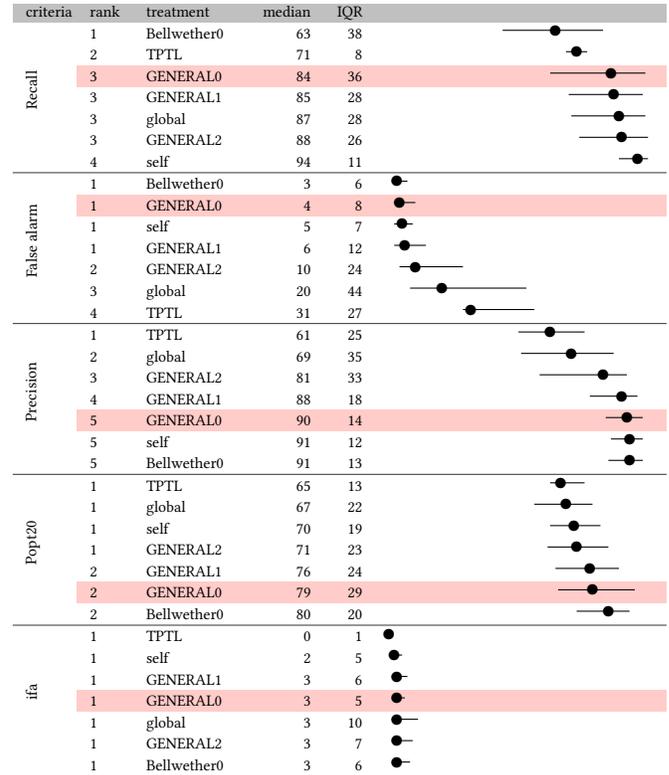
 
\footnotesize
\begin{tabular}{p{1.5cm}lp{1.2cm}rr}
\arrayrulecolor{darkgray}
\rowcolor[HTML]{C0C0C0} criteria & rank & treatment & median & IQR \\ 
 \multirow{7}{*}{\rotatebox[origin=c]{0}{Recall}} 
 &   1 &      Bellwether0 &    63 &  38  \\
 &   2 &      TPTL &    71 &  8  \\
 &   \cellcolor{lightpink}3 &      \cellcolor{lightpink}STABILIZER0 &    \cellcolor{lightpink}84 &  \cellcolor{lightpink}36  \\
 &   3 &      STABILIZER1 &    85 &  28 \\
 &   3 &      global &    87 &  28  \\
 &   3 &      STABILIZER2 &    88 &  26  \\
 &   4 &      self &    94 &  11 \\ \hline
\multirow{7}{*}{\rotatebox[origin=c]{0}{False alarm}} 
 &   1 &      Bellwether0 &    3 &  6  \\
 &   \cellcolor{lightpink}1 &      \cellcolor{lightpink}STABILIZER0 &    \cellcolor{lightpink}4 &  \cellcolor{lightpink}8  \\
 &   1 &      self &    5 &  7  \\
 &   1 &      STABILIZER1 &    6 &  12  \\
 &   2 &      STABILIZER2 &    10 &  24  \\
 &   3 &      global &    20 &  44  \\
 &   4 &      TPTL &    31 &  27  \\ \hline
\multirow{7}{*}{\rotatebox[origin=c]{0}{Precision}} 
 &   1 &      TPTL &    61 &  25  \\
 &   2 &      global &    69 &  35  \\
 &   3 &      STABILIZER2 &    81 &  33  \\
 &   4 &      STABILIZER1 &    88 &  18 \\
 &   \cellcolor{lightpink}5 &      \cellcolor{lightpink}STABILIZER0 &    \cellcolor{lightpink}90 &  \cellcolor{lightpink}14  \\
 &   5 &      self &    91 &  12  \\
 &   5 &      Bellwether0 &    91 &  13  \\ \hline
\multirow{7}{*}{\rotatebox[origin=c]{0}{Popt20}} 
 &   1 &      TPTL &    65 &  13  \\
 &   1 &      global &    67 &  22  \\
 &   1 &      self &    70 &  19 \\
 &   1 &      STABILIZER2 &    71 &  23  \\
 &   2 &      STABILIZER1 &    76 &  24  \\
 &   \cellcolor{lightpink}2 &      \cellcolor{lightpink}STABILIZER0 &    \cellcolor{lightpink}79 &  \cellcolor{lightpink}29  \\
 &   2 &      Bellwether0 &    80 &  20 \\ \hline
\multirow{7}{*}{\rotatebox[origin=c]{0}{ifa}} 
 &   1 &      TPTL &    0 &  1  \\
 &   1 &      self &    2 &  5  \\
 &   1 &      STABILIZER1 &    3&  6  \\
 &   \cellcolor{lightpink}1 &      \cellcolor{lightpink}STABILIZER0 &    \cellcolor{lightpink}3 &  \cellcolor{lightpink}5 \\
 &   1 &      global &    3 &  10  \\
 &   1 &      STABILIZER2 &    3 &  7 \\
 &   1 &      Bellwether0 &    3 &  6  \\
\end{tabular}
\caption{RQ3 results (for  defect prediction case study). The ``rank'' column at left comes from the statistical analysis methods of \S\ref{eval}. In this study, since  we want to minimize {\em false alarm} and {\em ifa}, then the {\em lowest ranks} (i.e., rank=1) are   {\em best}. But for other measures, which we want to maximize (recall, precision, Popt20), the {\em largest ranks} are {\em best}. 
}
\label{fig:Statistical_dp}
\end{figure}

\noindent
\textbf{(a)~Defect prediction:}
Figure~\ref{fig:Statistical_dp} shows the performance of all learners from  \S\ref{sec:mylearners} on the defect prediction case study. These results are grouped by the ``rank'' in the left-hand-side column (and this rank was generated using the statistical methods of \S\ref{eval}).

Looking at our results, in  Figure~\ref{fig:Statistical_dp}, we recommend STABILIZER0; i.e., use the single model found at the root of the STABILIZER tree. We say this since, in all performance measures, STABILIZER0 performs statistically similar or better than our baseline and SOTA methods while also returning the fewest models. To defend this recommendation, we note that:
\bi
    \item We would not recommend Krishna et al.'s Bellwether0 since its run-time is much slower $\Theta(N^2)$ and that longer run-time is not associated with an outstandingly  better performance.
    \item Nor would we recommend TPTL since it has the outstandingly worst false alarm rate and precision).
    \item Nor would we recommend STABILIZER1/STABILIZER2 since these never perform statistically different and better than STABILIZER0.  Also, they produce more models than STABILIZER0.
    \item Finally, we see our recommended model STABILIZER0 performs similar to self (which is reference model) in most cases, the only case where self beats STABILIZER0 is recall. This shows that our recommended model performs similarly to the reference model, indicating our transfer learning approach is reaching respectable performance while being able to stable conclusions.
    % \item Finally, we would not recommend ``self'' since there is no generalization there (it generates 756 models), and  the only case where it beats STABILIZER0 is recall. At the same time,  STABILIZER0 still offers respectable performance of 84\% recall.
\ei

Overall, we summarize Figure~\ref{fig:Statistical_dp} results as follows: ``Across multiple performance criteria, the faster STABILIZER0 never performs worse than Bellwether0. Furthermore, measured in terms of recall, STABILIZER0 performs statistically better. And thus suitable for transfer learning and generating stable conclusions''.

\begin{figure} 
\scriptsize
\begin{tabular}{p{.2cm}crrr|ccrrr}
\rowcolor[HTML]{C0C0C0} 
Goal                  & Rank & treatment       & med & IQR   & Goal                  & Rank & treatment       & med & IQR  \\ \hline
                      & \cellcolor{lightpink}1    & \cellcolor{lightpink}STABILIZER1 & \cellcolor{lightpink}0      & \cellcolor{lightpink}0.82  &                       & 1    & self            & 0.31   & 0.57 \\
                      & 1    & self            & 0.25   & 0.7   &                       & \cellcolor{lightpink}2    &\cellcolor{lightpink}STABILIZER1 & \cellcolor{lightpink}0.39   & \cellcolor{lightpink}0.79 \\
                      & 1    & STABILIZER2 & 0.25   & 0.85  &                       & 2    & STABILIZER0 & 0.46   & 1    \\
                      & 2    & STABILIZER0 & 0.31   & 1     &                       & 2    & STABILIZER2 & 0.53   & 0.93 \\
                      & 2    & Bellwether0     & 0.79   & 0.93  &                       & 2    & global          & 0.7    & 0.51 \\
\multirow{-6}{*}{\rotatebox[origin=c]{90}{MCP}} & 3    & global          & 0.9    & 0.5   & \multirow{-6}{*}{\rotatebox[origin=c]{90}{MOP}} & 2    & Bellwether0     & 0.77   & 0.68 \\ \hline
                      & 1    & self            & 0.37   & 0.44  &                       & 1    & self            & 0.67   & 0.57 \\
                      & 2    & STABILIZER2 & 0.52   & 0.44  &                       & 2    & STABILIZER2 & 0.76   & 0.53 \\
                      & \cellcolor{lightpink}2    & \cellcolor{lightpink}STABILIZER1 & \cellcolor{lightpink}0.56   & \cellcolor{lightpink}0.44  &                       & \cellcolor{lightpink}2    & \cellcolor{lightpink}STABILIZER1 & \cellcolor{lightpink}0.79   & \cellcolor{lightpink}1.13 \\
                      & 3    & Bellwether0     & 0.61   & 0.81  &                       & 2    & global          & 0.9    & 0.27 \\
                      & 3    & STABILIZER0 & 0.75   & 0.82  &                       & 2    & STABILIZER0 & 0.91   & 1.58 \\
\multirow{-6}{*}{\rotatebox[origin=c]{90}{MCI}} & 3    & global          & 0.75   & 0.82  & \multirow{-6}{*}{\rotatebox[origin=c]{90}{MOI}} & 2    & Bellwether0     & 1      & 0.5  \\ \hline
                      & 1    & self            & 0.54   & 1.09  &                       & 1    & self            & 0.38   & 0.41 \\
                      & \cellcolor{lightpink}1    & \cellcolor{lightpink}STABILIZER1 & \cellcolor{lightpink}0.6    & \cellcolor{lightpink}0.68  &                       & 2    & STABILIZER2 & 0.56   & 0.53 \\
                      & 1    & STABILIZER2 & 0.68   & 0.63  &                       & \cellcolor{lightpink}2    & \cellcolor{lightpink}STABILIZER1 & \cellcolor{lightpink}0.56   & \cellcolor{lightpink}0.41 \\
                      & 1    & STABILIZER0 & 1.07   & 2.04  &                       & 2    & STABILIZER0 & 0.65   & 0.48 \\
                      & 1    & global          & 1.16   & 1.79  &                       & 2    & global          & 0.65   & 0.89 \\
\multirow{-6}{*}{\rotatebox[origin=c]{90}{MC}}  & 2    & Bellwether0     & 8.75   & 59.48 & \multirow{-6}{*}{\rotatebox[origin=c]{90}{MS}}  & 3    & Bellwether0     & 2.94   & 5.43 \\ \hline
                      & 1    & self            & 0.25   & 0.43  &                       &      &                 &        &      \\
                      & 2    & STABILIZER2 & 0.38   & 0.49  &                       &      &                 &        &      \\
                      & \cellcolor{lightpink}2    & \cellcolor{lightpink}STABILIZER1 & \cellcolor{lightpink}0.39   & \cellcolor{lightpink}0.6   &                       &      &                 &        &      \\
                      & 2    & STABILIZER0 & 0.48   & 0.51  &                       &      &                 &        &      \\
                      & 2    & Bellwether0     & 0.5    & 0.74  &                       &      &                 &        &      \\
\multirow{-6}{*}{\rotatebox[origin=c]{90}{MAC}} & 2    & global          & 0.62   & 0.61  &                       &      &                 &        &     
\end{tabular}
\caption{RQ3 results (for project health).
Same format as
Figure~\ref{fig:Statistical_dp}.
For definitions
of our goal (MCP, MCI, MC. MAC, MS, MOI, MOP),
see \tbl{project_health_data}.}
\label{fig:Statistical_ph}
\end{figure}

\noindent
\textbf{(b)~Project health estimation:}
Figure~\ref{fig:Statistical_ph} shows the results of the project health estimation case study. Here the performance criterion is MRE, as mentioned in \S\ref{sec:Measures}. Unlike defect prediction case study, project health can be estimated in many ways, as shown in Table~\ref{tbl:feature_ph}. For each of the seven goals, the new algorithm STABILIZER is compared against the reference model (i.e. ``self'', that uses the local data), global model (``global''), and conventional bellwether (``Bellwether0'')

% \footnote{Now, one difference from defect prediction case study is that we do not compare against TPTL as it is a transfer learning approach designed explicitly for defect prediction.}.

% In all cases, ``self'' performs best in Figure~\ref{fig:Statistical_ph}. However, recall that this method offers no generalization (since it generates 1628 models). 

In all cases of project health estimation goals, our recommended STABILIZER1 did not beat the reference model. Our recommended model reached the reference model performance in two out of seven goals. At the same time, when compared to models that can generate stable conclusions, STABILIZER1 is statistically as  good as or better than Bellwether0, global, STABILIZER0 and STABILIZER2. Hence, if the goal is to generate stable conclusions, then the best method seen here (for project health estimation) is STABILIZER1.

% everything else, of the methods that generalize (i.e., return fewer models than the total number of projects), STABILIZER1 is statistically as  good as or better than Bellwether0, global, STABILIZER0 and STABILIZER2. Hence, if the goal is to generalize, then the best generalization method seen here (for project health estimation) is STABILIZER1.

% As to everything else, of the methods that generalize (i.e., return fewer models than the total number of projects), STABILIZER1 is statistically as  good as or better than Bellwether0, global, STABILIZER0 and STABILIZER2.

\subsection{RQ4:Is STABILIZER parsimonious (only generates a few models)?}
\label{sec:rq4}
To answer this question, we count the number of models generated at the best  level of STABILIZER's cluster tree.

According to \fig{Statistical_dp}, for defect prediction, the best level is the root of the tree (see the {\em STABILIZER0} results). Since the number of models returned here is one,  we can say that, for defect prediction, STABILIZER is very parsimonious.

According to \fig{Statistical_ph}, for project health estimation, the best level is the one just down from the root (see the {\em STABILIZER1} results). \fig{Statistical_ph} shows results from seven metrics, and there are some differences in the  models generated for each goal.  The size of the level1 trees seen in \fig{Statistical_ph} has a mode of nine models:

\begin{center}
    \textit{(MC:8, MOI:8, MAC:9, MOP:9, MCP:9,  MCI:9, MS:10)}
\end{center}

Note that nine projects are still far less than 1628 used in this study. Hence, we say that STABILIZER for project health is far more parsimonious than using raw project data.
  That said, STABILIZER for project health is  {\em less} parsimonious  than defect prediction (where only one model is found).

\subsection{RQ5:What are the stable conclusions we learned?}
\label{sec:rq5}
In this research question, we explore models which are stable across multiple projects and extract features from models from each case study. We do this to learn what metrics seemed to be most important for prediction and what conclusions can be generated from them? This is done by extracting features which are important for prediction according to the ML model built (i.e., impurity-based feature importance in our case extracted using the scikit-learn API). To determine which features are most important for prediction, we analyze both ``locally important'' and ``globally important'' features reported by the respective learners used in this study~\cite{pedregosa2011scikit}). Some features are ``globally important'' since our analysis shows that, across many projects, they are the ones that most predict for defects and estimate for project health. While ``locally important'' features are feature importance aggregated over all the training projects. We call these features  ``locally important'' since these results come from an analysis that is restricted to just one project at a time.

To show the feature importance of ``globally important'' and ``locally important'' features, we use two numbers $x/y$. Here the first number ($x$) is the aggregated feature importance for the {\em best STABILIZER method}  and the second number ($y$) is feature importance from the local model (self) aggregated over all training projects. Here ``globally important'' features are those where $x$ is large. Here, (a)~if $x/y$ is near one and both $x$ and $y$ are large, then that feature is both ``globally'' and ``locally'' important, (b)~if $x/y$ is near one and both $x$ and $y$ are small, then that feature is both ``globally'' and ``locally'' unimportant, (c)~if $x$ > $y$, then that feature is ``globally important'' but ``locally unimportant'', and (d)~if $x$ < $y$, then that feature is ``locally important'' but ``globally unimportant''.

\subsubsection{Defect Prediction}
Table~\ref{tbl:feature_dp} shows the ``globally important'' features (marked in \colorbox{lightpink}{pink}) in the STABILIZER0 defect prediction models. We observe that only three out of 21 features are important in predicting defects. Among these three features, (a) two are experience based metrics; (b) one is change related metric. We analyzed the decisions made by our model to differentiate between a defective and non-defective entry to generate recommendations for the community of 756 Github projects studied here\footnote{Note: the features which are  important in predicting defects are based on the analysis of 756 studied here. So the conclusions generated are applicable for these projects only.}  

\bi
    \item The presence of {\em age} in the model suggests: {\bf review codes which are changed more often. The frequently a file is changed, the more problematic it can become.}
    \item  The presence of {\em exp} (experience) in our models lets us make the recommendation: {\bf minimize the use of amateurs dabbling around the codebase in areas that are unfamiliar to them}.
    \item {\em rexp} (recent developer experience) as an important feature suggests: {\bf review codes more thoroughly when the changes are made by developers not familiar with the latest changes}.
\ei

\begin{table}[!t]
\footnotesize
\caption{Distribution of feature importance using the self model and the STABILIZER0 (Defect Prediction). The \colorbox{lightpink}{pink} cells show important features from STABILIZER0. Each cell consists of two numbers(like x/y): the first one (x) is the feature importance for the STABILIZER0 model, and the second one (y) is feature importance for the self model. Learner is Random Forest.}
\label{tbl:feature_dp}
\begin{tabular}{l|l|ll}
\rowcolor[HTML]{C0C0C0} 
Feature      & Feature Importance               & Feature                                & Feature Importance                \\ \hline
la    & 0.0/0.02                         & \multicolumn{1}{l|}{avg\_nddev} & 0.07/0.04                         \\
ld    & 0.0/0.05                         & \multicolumn{1}{l|}{avg\_nadev} & 0.0/0.03                          \\
lt    & 0.0/0.01                         & \multicolumn{1}{l|}{avg\_ncomm} & 0.0/0.01                          \\
age   & \cellcolor{lightpink}0.22/0.1 & \multicolumn{1}{l|}{ns}         & 0.0/0.0                           \\
ddev  & 0.0/0.09                         & \multicolumn{1}{l|}{exp}        & \cellcolor{lightpink}0.21/0.03 \\
nuc   & 0.0/0.09                         & \multicolumn{1}{l|}{sexp}       & 0.0/0.02                          \\
own         & 0.09/0.07                        & \multicolumn{1}{l|}{rexp}       & \cellcolor{lightpink}0.38/0.07 \\
minor       & 0.02/0.15                        & \multicolumn{1}{l|}{nd}         & 0.0/0.0                           \\
ndev  & 0.0/0.05                         & \multicolumn{1}{l|}{sctr}       & 0.0/0.07                          \\
ncomm & 0.0/0.02                         & \multicolumn{1}{l|}{nadev}      & 0.0/0.01                          \\
adev  & 0.0/0.04                         & \multicolumn{1}{l|}{}                 &                                  
\end{tabular}
 
\end{table}

\begin{table}[]
\scriptsize
% \caption{Distribution of features importance using the self model and the STABILIZER1 model (project health). Each column represents one of the goals mentioned in Table~\ref{tbl:project_health_data}.  The \colorbox{lightpink}{pink} cells show important features from STABILIZER1 derived from the underlying model (i.e., Decision Tree Regressor). Each cell consists of two numbers(like x/y): the first one (x) is the feature importance for the STABILIZER1 model and the second one (y) is feature importance for the self model.}
\caption{Distribution of features importance using the self model and the STABILIZER1 model (project health). Each column represents one of the goals mentioned in Table~\ref{tbl:project_health_data}.  The \colorbox{lightpink}{pink} cells show important features from STABILIZER1 derived from the underlying model (i.e., Decision Tree Regressor). Each cell consists of two numbers(like x/y): the first one (x) is for the STABILIZER1 model and the second one (y) is for the self model.}
\label{tbl:feature_ph}
\begin{tabular}{cc|ccccccc}
\rowcolor[HTML]{C0C0C0} 
\multicolumn{2}{c}{\cellcolor[HTML]{C0C0C0}}                   & \multicolumn{7}{c|}{\cellcolor[HTML]{C0C0C0}Target Variables}                                                                                                                                                                                       \\
\rowcolor[HTML]{C0C0C0} 
\multicolumn{2}{c}{\multirow{-2}{*}{\cellcolor[HTML]{C0C0C0}}} & MC                                & MAC                               & MOP                            & MCP                               & MOI                               & MCI                            & MS                                \\ \hline
                                                & MCI          & 0/0.11                            & 0.02/0.11                         & 0/0.1                          & \cellcolor{lightpink}0.17/0.13 & 0/0.07                            & NA                             & 0/0.08                            \\
                                                & MCP          & \cellcolor{lightpink}0.46/0.06 & 0/0.02                            & 0/0.08                         & NA                                & 0/0.02                            & 0/0.08                         & 0/0.05                            \\
                                                & MC           & NA                                & \cellcolor{lightpink}0.14/0.07 & 0/0.09                         & 0.01/0.09                         & 0/0.09                            & 0/0.05                         & 0/0.1                             \\
                                                & MAC          & \cellcolor{lightpink}0.28/0.09 & NA                                & 0/0.07                         & 0/0.04                            & 0/0.04                            & 0/0.08                         & 0/0.07                            \\
                                                & MF           & 0/0.08                            & \cellcolor{lightpink}0.61/0.08 & 0/0.06                         & 0/0.07                            & 0/0.08                            & 0/0.08                         & 0/0.24                            \\
                                                & MIC          & 0/0.1                             & 0.01/0.1                          & 0/0.07                         & 0/0.08                            & \cellcolor{lightpink}0.22/0.09 & \cellcolor{lightpink}1/0.09 & 0/0.13                            \\
                                                & MMP          & 0.01/0.1                          & 0/0.1                             & \cellcolor{lightpink}1/0.19 & 0/0.12                            & 0/0.08                            & 0/0.1                          & \cellcolor{lightpink}0.13/0.09 \\
                                                & MOI          & 0/0.07                            & 0.04/0.03                         & 0/0.03                         & 0/0.08                            & NA                                & 0/0.08                         & 0/0.04                            \\
                                                & MOP          & 0/0.11                            & 0/0.13                            & NA                             & 0.01/0.16                         & 0/0.06                            & 0/0.09                         & \cellcolor{lightpink}0.2/0.07  \\
                                                & MPC          & 0/0.08                            & \cellcolor{lightpink}0.17/0.08 & 0/0.11                         & \cellcolor{lightpink}0.81/0.09 & 0/0.11                            & 0/0.06                         & \cellcolor{lightpink}0.67/0.06 \\
                                                & MPM          & \cellcolor{lightpink}0.25/0.05 & 0/0.07                            & 0/0.06                         & 0/0.04                            & 0/0.08                            & 0/0.08                         & 0/0.02                            \\
                                                & MS           & 0/0.07                            & 0/0.09                            & 0/0.08                         & 0/0.07                            & \cellcolor{lightpink}0.78/0.18 & 0/0.1                          & NA                                \\
\multirow{-13}{*}{\rotatebox[origin=c]{90}{Predictor Variables}}          & MW           & 0/0.04                            & 0/0.09                            & 0/0.01                         & 0.01/0.02                         & 0/0.07                            & 0/0.09                         & 0/0.03                           
\end{tabular}
 \end{table}

\subsubsection{Project health estimation:}
Table~\ref{tbl:feature_ph} shows the features which are ``globally important'' (marked in \colorbox{lightpink}{pink}) for each of the seven project health estimation goals explored in this study. Each row in the table represents a predictor variable (13 variables as suggested by prior work (see \S~\ref{sec:data})), while each column represents one goal (seven goals in total). For each goal, we ask which features in practice are useful when we estimate project health. In this study, using the 1200  projects (from the group of 1628 projects) selected for training, we only find one to three features out of 13 are important.

In summary, we can say:

\bi
    \item \textit{MC}: for predicting the number of commits in the future, the important features are related to collaboration (MCP, MPM) and activeness (MAC) of the project. 
    \item \textit{MAC}: for predicting how many active contributors will be there, level of activeness(MC), project popularity (MF), and collaboration (MPC) play an essential role.
    \item \textit{MOP}: for estimating how many open PRs will still be active, the level of collaboration is a good indicator.
    \item \textit{MCP}: enhancement (MCI) and amount of collaboration (MPC) in a project are good indicators of how many PRs will be closed in the next six months.
    \item \textit{MOI}: how many new issues will be in open status in the next six months can be estimated from the amount of enhancement (MIC) and popularity (MS) in a project.
    \item \textit{MCI}: the amount of enhancement (MIC) is an accurate indicator for the number of issues closed in the future.
    \item \textit{MS}: for estimating the number of stargazers, the amount of collaboration (MMP, MOP, and MPC) is a good indicator.
\ei

\section{Discussion}

In this section we try to discuss what makes our STABILIZER framework unique and why it should be used based on the results of our study and goals set for this research. 

\textbf{Scalable:} Unlike the bellwether method proposed by Krishna et al., our STABILIZER framework is scalable. The results from RQ~\ref{sec:rq1} shows that STABILIZER framework is order of \textit{m times} faster than Bellwether0 (aka bellwether method proposed by Krishna et al.~\cite{krishna2018bellwethers}) in-case of theoretical average case computational complexity (expressed in-term of $\theta$), where $m$ is the number of clusters formed with $N$ projects. While the results from RQ~\ref{sec:rq2} shows that shows that in both case studies STABILIZER framework if much more scalable as number of projects increases, as can be seen in Figure~\ref{fig:compare} and \ref{fig:compare_time}.

\textbf{Largest Case Study:} Exploiting this scalability,
we have shown here that
  STABILIZER  with  
  two very large cases studies
(one with 756 projects for defect prediction case study and one with 1628 projects for project health estimation case study). In terms
of transfer learning lessons learned between multiple projects, this paper is the largest such case study yet reported in the software analytics literature. As stated above  in section~\ref{sec:intro},
most prior papers   reported their results based on  only 20 projects (median value).

\textbf{Model-agnostic:}
While version of  STABILIZER used here was only tested on random forests, we note that there is nothing in principle preventing the application of this method to a range of other learners. For example, when testing a large number of variants of a deep learner, we would predict that STABILIZER's divide-and-conqueror (and promote) approach  would speed up that analysis.

\textbf{Performance:} We can effectively use the STABILIZER framework if and only if it shows similar or better performance than reference and SOTA methods. In RQ~\ref{sec:rq3} we compared our STABILIZER framework with other SOTA and reference models in-term of prediction performance measured with 5 widely used metrics for defect prediction case study and 1 for project health estimation case study. We see that the STABILIZER reaches similar performance to the reference model while performing better then SOTA methods for defect prediction case study, while for project health estimation STABILIZER did not show similar performance to the reference model, but it did out performed other SOTA methods. This result signifies that STABILIZER framework performs significantly better, while being significantly faster. 

\textbf{Conclusion Stability:} One goal of this researcher is to find conclusions which are stable across a larger number of projects. If we can't effectively reduce the number of models generated then we had not solved the problem of conclusion instability. the results from RQ~\ref{sec:rq4} shows that can achieve conclusion stability using the STABILIZER framework as the framework found 1 project which can be used as bellwether project for the whole community, signifying the features learned from the model build from that project is stable across 756 projects for the defect prediction case study. While for the project health estimation case study, we find median of 9 projects each specialized for a set of $\approx 200$ projects, signifying features learned from each of these bellwether project's models is stable across those $\approx 200$ projects. This shows the STABILIZER framework can achieve conclusion stability across a large community of project. 

\textbf{Generally-used Conclusions:}  Conclusion stability says  that   new data is not perpetually changing old models. 
Once that is achieved, the next question to ask is ``are those stable conclusions useful?''.  RQ~\ref{sec:rq5} shows what conclusions can be reached about what features are important in predicting defects and estimating project health in the respective case studies and explains in details whats the importance of each feature in the stable conclusion.

\section{Threats to Validity}
\label{sec:validity}
As with any large-scale empirical study, biases can affect the final results. Therefore, any conclusions made from this work must be considered with the following issues in mind:

(a) \textit{Evaluation Bias}: 
While those results  are accurate, the conclusion is scoped by the evaluation metrics. It is possible that using other measurements; there may well be a difference in these different kinds of projects. This is a matter that needs to be explored in future research. 

(b) \textit{Construct Validity}: We made many engineering decisions using advice from the literature. That said,  we acknowledge that other constructs might lead to different conclusions. 

(c) \textit{External Validity}: The defect prediction dataset was collected using ``commit\_guru''. It is possible using other tools or methods may result in different outcomes. That said, the ``commit\_guru'' is a tool that is widely used and has detailed documentation. 

(d) \textit{Statistical Validity}: We applied two statistical tests, bootstrap, and the a12. Hence, anytime in this paper, we reported that ``X was different from Y'' then that report was based on both effect size and statistical significance test.
 
(e) \textit{Sampling Bias}: Our conclusions are based on the 756 projects (defect prediction) and 1628 projects (project health estimation). It is possible that different projects would have lead to different conclusions. That said, this sample is very large, so we have some confidence that this sample represents an interesting range of projects.

\section{Conclusion}
\label{sec:concl}
Software engineering suffers from the conclusion instability problem. Ideally, we should be able to generalize to a single model. However, given the diversity of SE, we warn that it is folly to assume that all projects can be represented within a single model.

We presented a method called STABILIZER that recursively replaces many models (clustered into a tree) with one model per sub-tree. That model is then moved up the tree to see if it can replace any higher-level models. Sometimes, this process results in a single model (as in the case of defect prediction). On the other hand, sometimes it does not: for project health estimation, we found that STABILIZER reports nine (median values) models per goal.

It should be emphasized that this approach has only been tested from predictions of defects or project health.  However, within that restriction, we say that it is not necessary to always reason about each new project as a new concept that might generate a new model. Even in our worse case (project health), 1628 projects can be represented as nine exemplary models (median value).

Looking forward, based on this conclusion, we make two comments. Firstly, we think this work offers a more rational basis for SE research and SE textbooks. SE is a diverse process,  but at least some aspects of that work can be captured in a handful of models. Such a small set of models could be the basis of (e.g.) a 14-week graduate class (one model per week) or a product line of software project tools (one product per model).  

Secondly, we warn that much of the prior work on homogeneous transfer learning may  have complicated the process with needlessly complicated methods. When building increasingly complex and expensive methods, we strongly recommend that researchers pause and compare their sophisticated methods against simpler alternatives.  Going forward from this paper, we would recommend that the transfer learning community uses STABILIZER as a baseline method against which they can test more complex methods.

% \section{Acknowledgements}
% This paper was partially funded by blinded for review.
% \balance
% bibliographystyle needs to be changed - look at MSR website

% \bibliographystyle{IEEEtran} % for arxiv 
\bibliographystyle{ACM-Reference-Format}
\bibliography{main}

\end{document}